\documentclass[11 pt]{article}
\usepackage{amsmath}
\usepackage{amssymb}
\usepackage{latexsym}
\usepackage{graphicx}
\usepackage{color}
\usepackage{caption}
\usepackage{subcaption}
\usepackage[colorlinks=true, citecolor=blue]{hyperref}
\usepackage{dcolumn}% Align table columns on decimal point
\usepackage{bm}% bold math

\usepackage{hyperref}

\def\beq{\begin{equation}}
\def\eeq{\end{equation}}
\def\be{\begin{equation}}
\def\ee{\end{equation}}
\def\ba{\begin{array}}
\def\ea{\end{array}}
\def\bea{\begin{eqnarray}}
\def\eea{\end{eqnarray}}

\def\eps{\epsilon}

\newcommand{\amp}{A(\theta)}
\newcommand{\ampk}{A(\epsilon k)}
\newcommand{\n}[1]{\hat{n}_{#1}}

\newcommand{\N}[1]{n_{#1}^{\mu}}

\def\a{\alpha}
\def\m{\mu}
\def\D{\Delta}
\def\d{\delta}
\def\e{\epsilon}
\def\o{\omega}
\def\s{\sigma}
\def\ra{\rangle}
\def\la{\langle}
\def\half{\frac12}

\begin{document}

\title{\bf Spin on a 4D Feynman Checkerboard}

\author{Brendan Z. Foster\footnote{\texttt{bzfoster@bzfoster.com}}~  
 and Ted Jacobson\footnote{\texttt{jacobson@umd.edu}, 
{\it Maryland Center for Fundamental Physics,
University of Maryland, College Park, MD 20742-4111}}}

% \email{bzf@physics.umd.edu}
%\altaffiliation[Present address: ]
%{Insitut d'Astrophysique de Paris, 98 bis Bvd.~Arago, 75006 Paris,
%FRANCE}%Lines break automatically or can be forced with \\
%\affiliation{Department of Physics, University of Maryland\\ College
%Park, MD 20742-4111 USA}%Lines break automatically or can be forced with\\
%\author{Ted Jacobson}%
% \email{jacobson@physics.umd.edu}
 %\altaffiliation[Present address: ]
%{Insitut d'Astrophysique de Paris, 98 bis Bvd.~Arago, 75006 Paris,
%FRANCE}%Lines break automatically or can be forced with \\
%\affiliation{Department of Physics, University of Maryland\\ College Park,
%MD 20742-4111 USA}

\date{ }

%\begin{abstract}

\maketitle
\abstract
We discretize the Weyl equation for a massless, 
spin-1/2 particle on a time-diagonal, hypercubic spacetime lattice
with null faces. The amplitude for a step of right-handed chirality
is proportional to the spin projection operator 
in the step direction, while for left-handed it is the orthogonal projector.
Iteration yields a path integral for the retarded propagator, 
with matrix path amplitude proportional to the product of 
projection operators. This assigns the amplitude $i^{\pm T}\, {3}^{-B/2}\,2^{-N}$ to a path 
with $N$ steps, $B$ bends, and $T$ right-handed minus left-handed bends, where the sign 
corresponds to the chirality.   
Fermion doubling does not occur in this discrete scheme. 
A Dirac mass $m$ introduces the amplitude $i\e m$ to flip chirality in any given time step $\e$,
and  a Majorana mass similarly introduces a charge conjugation amplitude.

%\end{abstract}


\section{Dedication}

To break the spacetime code was the aim of much of David Finkelstein's work \cite{Finkelstein:1970xm,Finkelstein:1996fn}.
David's approach was influenced, inter alia, by von Neumann's quantum logic, 
Feynman's operator calculus \cite{Feynman:1951gn}, and his own deep insight into
quantum mechanics as a physics of processes, not things.
He envisioned spacetime as an auto-generated algebra, a quantization of the natural
numbers, possessing enough structure and depth to serve as the cosmological 
process and to generate the local symmetries of particle physics \cite{Finkelstein:1989wi,
Finkelstein:1996wu}. 
He proposed that the continuum is a coherent quantum phenomenon \cite{Finkelstein:1989wi}:
\begin{quote}
\it 
Vacuum I, the present ambient space-time, with its Minkowski
chronometry, is a critical phenomenon, a Bose condensation of
chronon pairs into a hypercubical lattice of basic states.
\end{quote}
Though his vision did not guide him 
all the way to a complete new physics, he may well have been on the
true path. He was nearly alone in his quest, but we imagine that
others will eventually catch up, catch on, 
and carry it onward.
The present paper meets David on the time-diagonal hypercubical lattice. But while for him it represented a coherent
quantum process, for us it is but a classical scaffold on which the propagation of spin-1/2
particles plays out. We dedicate this paper to David's memory, with gratitude for his insight and inspiration.

\section{Introduction}

\begin{quote}
{\it 
``And, so I dreamed that if I were clever, I would find a formula for the amplitude of a path that was beautiful and simple for three dimensions of space and one of time, which would be equivalent to the Dirac equation, and for which the four components, matrices, and all those other mathematical funny things would come out as a simple consequence - I have never succeeded in that either."} 
\end{quote}
This quote is taken from Feynman's Nobel lecture \cite{Feynman:1965jda}, 
where it appears
just after he describes his sum over checkerboard paths for
the Dirac propagator in 1+1 dimensions (see also \cite{qmpi}).
A path is composed of steps on a square lattice of null links,
and its amplitude is $(i\epsilon m)^R$, where $m$ is the particle mass,
$R$ is the number of direction reversals, and  $\epsilon$ is the
time duration of a lattice step (and we use units
with $\hbar=c=1$). If the mass vanishes the particle moves only
to the left or to the right, at the speed of light, and these two
motions correspond to the two chiralities for a
Dirac spinor in 1+1 dimensions.

This checkerboard path integral intriguingly accounts for relativistic
propagation and Dirac matrices with nothing more than a 
simple factor of $i$ associated with a geometric 
property---a bend---in a piecewise lightlike
path. Here we show that
one can actually come very close to preserving these 
striking features in a path integral for Dirac 
particles in $3+1$ dimensions.

Most work on lattice
formulations of spinor propagation
has been directed at lattice field theory calculations,
and thus involves Grassmann variables and 
``path" integrals over field configurations
in a spacetime of Euclidean signature \cite{x}.
Here we pursue instead a
bosonic (i.e.~non-Grassmanian)
formulation in terms of a sum over particle paths
on a Minkowski signature lattice.
Such formulations have been studied 
previously in~\cite{Riazanov,TJdiss,Bialynicki-Birula:hi} and
references therein. 
What is new here is the
interesting structure of the lattice employed, and the
fact that it allows for a particularly simple rule
for the amplitudes. 

The massless case 
in 3+1 dimensions is far more interesting
than in 1+1 dimensions.
Chiral, right-handed 
two-component spinors satisfy the 
Weyl equation,
\beq
\sigma^\mu\partial_\mu\Psi=0,
\label{covWeyl}
\eeq
where $\sigma^{\mu}=(1,\vec{\sigma})$
with $\vec{\sigma}$ the Pauli matrices, and
left handed spinors satisfy the analogous 
equation with $\bar\sigma^{\mu}=(1,-\vec{\sigma})$.
As in 1+1 dimensions, the mass introduces an amplitude
$i\epsilon m$ to flip chirality at each step. 
The key point is that 
the velocity operator is $\pm$ the 
spin operator for a right- or left-handed Weyl particle.
As we shall see, this implies that 
displacement amplitudes are determined
by spin transition amplitudes.
 
\section{Lattice} 
 
\begin{figure}[t]
\vbox{% \vskip 8 pt
\centerline{\includegraphics[width=5cm]{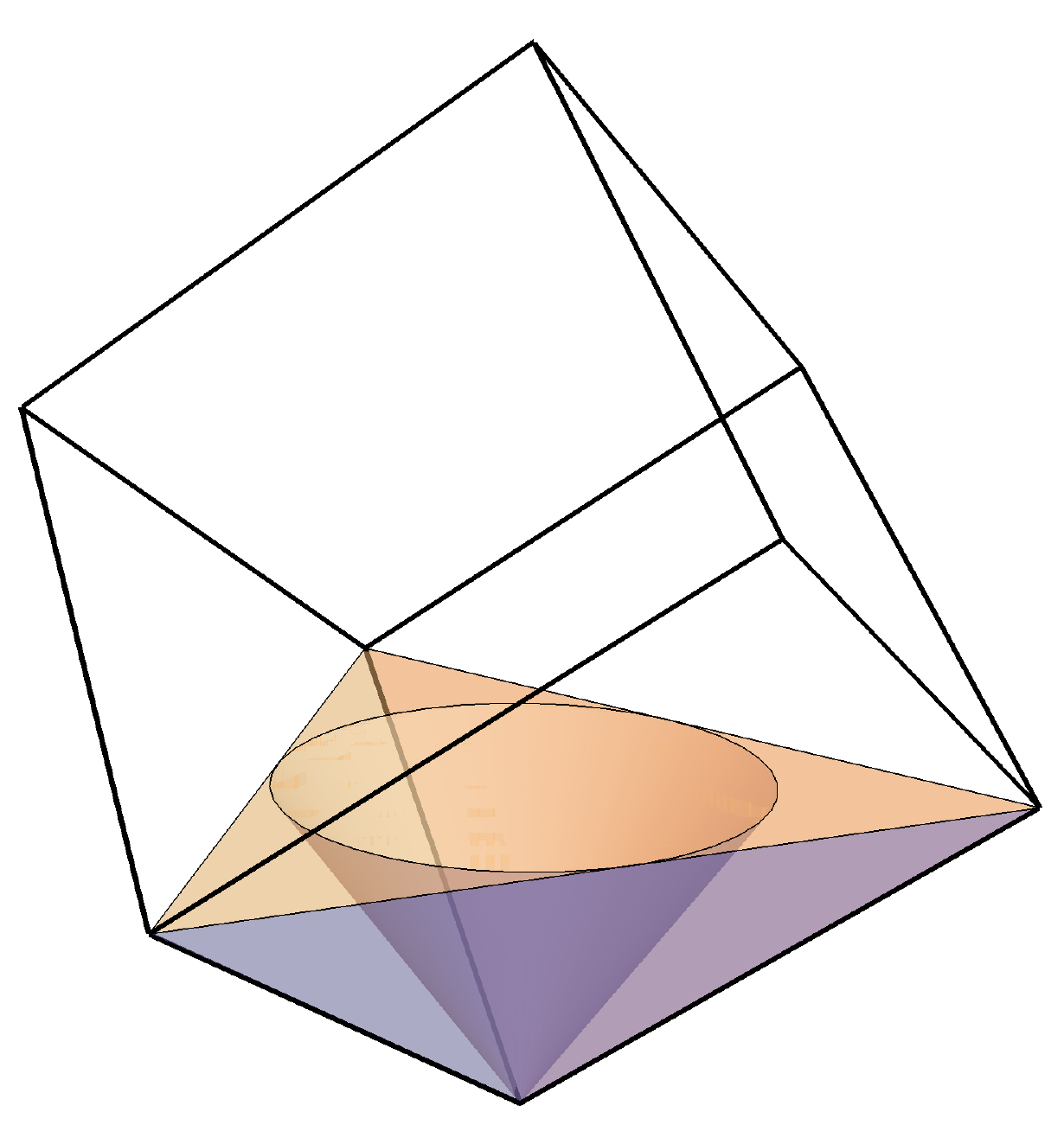}}
\caption{\label{cone}
\small Unit cell of a time-diagonal cubic spacetime lattice in $2+1$ dimensions. The cube faces are null planes, so that the continuum lightcone nestles tangent to the cube, and the edges are spacelike with speed $2$ relative to the diagonal.
\smallskip}
}\end{figure}

We discretize spacetime with a
hypercubical lattice, oriented so one
diagonal of the hypercube lies in the time
direction, and with the step speed chosen
three times the speed of light,
so that the discrete causal cone just encloses
the continuum one (see Fig.~\ref{cone}).
The edges from one
vertex lie in the directions of the
four spacetime vectors
\beq
\N{i}=(1,\a \n{i}).
\label{Ni}
\eeq
The four spatial unit vectors $\n{i}$ point to
the vertices of a tetrahedron, and $\a$ is the step
speed. 
The $\n{i}$ sum to zero, and the inner product
or angle between any distinct two is the same.
Hence for $i\ne j$ we have
$\n{i}\cdot\n{j}=-1/3$, and the angle is
equal to $\cos^{-1}(-1/3) \approx 109^\circ$.
An explicit expression for the components
of the $\n{i}$ in
a particular basis is
\bea\label{COORD}
\n{1}&=&\textstyle{\frac{1}{3}}(2\sqrt{2},0,-1)\nonumber\\
\n{2}&=&\textstyle{\frac{1}{3}}(-\sqrt{2},\sqrt{6},-1)\nonumber\\
\n{3}&=&\textstyle{\frac{1}{3}}(-\sqrt{2},-\sqrt{6},-1)\nonumber\\
\n{4}&=&(0,0,1).
\eea

It might seem natural to choose the step speed
$\a=1$, so that the links $\N{i}$ of the lattice would
be null, as envisaged in~\cite{Finkelstein:1996wu},
and as implemented in a lattice formulation of general relativity  
in \cite{Schaden:2015wia}.
However, in this case our retarded
lattice propagator would fail to converge at all in
the continuum limit. The reason is that such a spacetime
lattice violates the well-known ``Courant condition" for stability:
the discrete region of causal influence 
must contain the continuum one.

To marginally satisfy the Courant condition,
the tetrahedral cone, formed by the
four hyperplanes spanned by three of the
$\N{i}$'s, must barely enclose
the spherical, continuum light cone.
That is, the faces of the hypercube must be null. 
Equivalently, the dual lattice must be generated
by null covectors.\footnote{See \cite{Neiman:2012fu} for a general investigation
of spacetime polytopes with null faces.} 
Each of these hyperplanes
must therefore contain one and only one
null direction. By symmetry this null direction $k^\mu$ must
coincide with the sum of the three link vectors, e.g.
$k^\mu=\bigl(3,\a(\n{1}+\n{2}+\n{3})\bigr)$. The Minkowski
norm of this vector is $9-\a^2$, hence 
we must choose $\a=3$ if it is to
be null. Moreover,
$k^\mu n_{1\mu}=3-(\a^2/3)$, so if $\a=3$ the null
vector $k^\mu$
is orthogonal to all vectors in the hyperplane,
confirming that the hyperplane is indeed null.

\begin{figure}[t]
\vbox{% \vskip 8 pt
\centerline{\includegraphics[width=4cm]{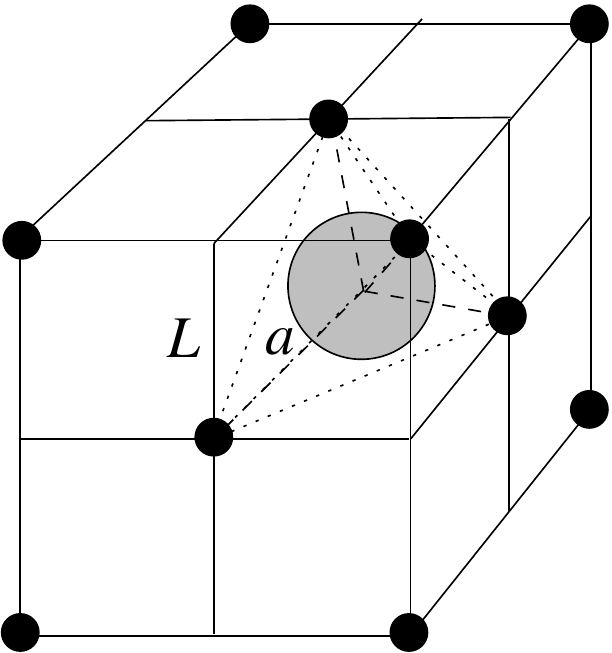}}
\caption{ \label{fcc}
\small Face-centered cubic lattice of points at one time
step. The tetrahedron (dotted lines)
is comprised of the four points reached
from the center of the small cube in one time step
(dashed lines).
The continuum sphere of light is enclosed by and tangent to the
tetrahedron. The distance from the center to a
tetrahedron vertex is three times the radius of the sphere.
The step length $a$ and cube edge length $L $ are shown.
\smallskip}
}\end{figure}

The spatial lattice at one time is a face-centered
cubic (fcc) lattice. A way to see this is to begin with
the tetrahedron of points that lies at
one time step to the future of a given spacetime point $p$. 
The four dimensional lattice has translation
symmetries that map any point to any other point,
and the spatial lattice at one time must share this property.
Hence it can be grown from this tetrahedral seed
by translation along the edges of the tetrahedron,
which produces the fcc lattice shown in Fig.~\ref{fcc}.

Evolving the spatial lattice one time step to the future
amounts to shifting it along the displacement from
the center of one tetrahedron to one of its vertices,
yielding a distinct but equivalent spatial lattice.
After four such steps the original spatial
lattice is recovered.
One can visualize these steps all in the same
direction, each one extending for half a diagonal
of one of the constituent cubes in Fig.~\ref{fcc}.
Alternatively, one can think of them as successively along
four distinct edges of the hypercube, which makes it
obvious that four steps brings one back to the original
lattice.

It will be important later to know the spatial volume per point
at one time step, expressed in terms of the step length.
It is clear from Fig.~\ref{fcc} that each point can be
associated uniquely with a pair of constituent
cubes of edge length $L $.
The volume of these cubes is $2L ^3$. The step
length $a$ is half the diagonal of one of these cubes,
hence $a=\sqrt{3}L /2$, so the volume per point
$V_p$ is given by
\beq
V_p= \frac{16}{3\sqrt{3}}a^3.
\label{Vp}
\eeq
We will take the time step size to be $\epsilon$, so the spatial step size is 
$a=\a \epsilon$.

\section{Discrete Weyl equation}
Now consider the tetrahedral quartet of unit vectors
$\n{i}$ defined below \eqref{Ni}. The sums $\sum_i \n{i}^a$ and
$\sum_i\n{i}^{a}\n{i}^{b}$
(with `$\n{i}^{a}$' denoting the `a'component of $\n{i}$)
are invariant under the symmetries of the tetrahedron,
hence the first sum must vanish and the second sum
must be proportional to the Euclidean metric
$\delta^{ab}$. Since the trace is equal to four this yields
the relation
\be
\label{BASIS}
\sum_i\n{i}^{a}\n{i}^{b}=\frac{4}{3}\delta^{ab}.
\ee
Using these identities and the
definition (\ref{Ni}) of the 4-vectors $\N{i}$,
the matrix 4-vector $\sigma^{\mu}=(1,\vec{\sigma})$ can
be expressed as
\be
\sigma^{\mu}=\frac{1}{2}\sum_{i}
\frac{1}{2}\left(1+\frac{3}{\alpha}\n{i}\cdot\vec{\sigma}\right)\N{i}.
\label{sigmaalpha}
\ee
In the special case $\a=3$, for which the polyhedral light
cone conists of null hyperplanes, this becomes just
\be
\sigma^{\mu}=\frac{1}{2}\sum_{i}P_{i}\, \N{i},
\label{sigma}
\ee
where 
\be
P_i={\textstyle\frac{1}{2}}(1+\n{i}\cdot\vec{\sigma})
\label{Projector}
\ee
is the projector for spin up in the direction $\n{i}$.
Note in particular that $\half \sum_i P_i=I$.
(A version of this construction 
involving an integral over the sphere of directions 
was developed in \cite{Jacobson:xt}.)

Using the identity (\ref{sigma}),
the Weyl equation (\ref{covWeyl}) for right-handed
spinors takes the form
\beq
\frac{1}{2}\sum_{i}P_{i}\N{i}\partial_\mu\Psi=0.
\label{PWeyl}
\eeq
If we approximate the directional derivatives
by finite differences,
\be
\epsilon \N{i}\partial_{\mu}\Psi(x)
\cong\Psi(x)-\Psi(x-\epsilon n_{i}),
\label{difference}
\ee
(\ref{PWeyl})
yields a formula 
for $\Psi(x)$ on the lattice in terms of
the values $\Psi(x-\epsilon n_{i})$
at the immediately preceding points,
\be
\Psi(x)=\frac{1}{2}\sum_{i}P_{i}\Psi(x-\epsilon n_{i}).
\label{onestep}
\ee
This is our one-step evolution rule. It states that a
prior amplitude contributes to $\Psi(x)$ precisely to the 
extent that its spin is parallel to the spatial step 
arriving at $x$. 

\subsection{Dispersion relation}
We can find the lattice dispersion relation by considering a wave function
with plane wave form,    
\beq\label{planewave}
\Psi(x,0) = e^{-i k_\mu x^\mu} \Psi_0,
\eeq
where $\Psi_0$ is a constant spinor.
This satisfies \eqref{onestep} if and only if 
\beq
\Psi_0 = \frac{1}{2}\sum_j P_j  e^{i\eps k_\mu n_j^\mu} \Psi_0.
\label{nodoubles}
\eeq
For small $\eps k$, we may expand to first order,
and using equation (\ref{sigma})
we find the standard
Weyl equation for momentum eigenstates, 
$\sigma^\mu k_\mu \Psi_0=0$.
These solutions obey the relativistic
dispersion relation, $k_\mu k^\mu=0$, at $O(\epsilon)$.
For generic $\eps k$, \eqref{nodoubles} imposes the condition 
that the matrix multiplying $\Psi_0$ must have
a unit eigenvalue, $\Psi_0$ being the corresponding eigenvector. 
This condition is the exact dispersion relation. 

To extract its physical 
meaning, we assume that the spatial components $k_a$ of 
$k_\m$ are real, and solve for the frequency $\o=k_0$, which in general will be complex. 
First we recast \eqref{nodoubles} in the form 
\beq
A(\theta) \Psi_0 = e^{-i\o\e}\,\Psi_0, 
\label{nodoubles2}
\eeq
with
\beq
\amp=\frac{1}{2}\sum_jP_j\, e^{i\theta_j},
\label{A}
\eeq
and  
$\theta_j:=\eps k_a n^a_j$. 
Since $\sum_j n^a_j=0$, we also have $\sum_j\theta_j=0$.
So the problem is to find the eigenvalues of $A(\theta)$, when the 
four $\theta_j$'s sum to zero.
Let us begin by finding when the frequency $\o$ is real, i.e.\
when the eigenvalue has unit modulus.

It is shown in the Appendix that the 
eigenvalues of $\amp$ have less than unit modulus except when at least three of the
$\theta_j$ coincide.\footnote{The eigenvalues 
of $\amp$ have modulus less than or equal to unity if and only if the step speed
$\a$ is $\ge3$. Our choice $\a=3$ is thus the 
marginal value for a convergent finite difference scheme.} 
Hence it suffices to consider the case  
when $\theta_1=\theta_2=\theta_3$, and $\theta_4=-3\theta_1$.
Then 
$\amp$ takes the form
\beq\label{case}
\amp=\cos 2\theta_1\, e^{-i\theta_1} P_4 + e^{i\theta_1} \tilde{P}_4,
\eeq
where $\tilde{P}_4= 1-P_4$ is the projector orthogonal to $P_4$.
The only nontrivial eigenvector solution with real frequency thus has
spin in the direction $-\hat{n}_4$,  
and frequency $\o=-\theta_1/\e$. The wavevector generally
satisfies $k_\mu n_{j}^\mu=\o + k_a n^a_{j} = (-\theta_1+\theta_j)/\e$.
For this solution we thus have $k_\mu n_{1,2,3}^\mu=0$, which implies that
$k_\mu$ is precisely the null normal to the $123$ (null) hyperplane.
The only solutions to the lattice dispersion relation with real frequency 
are thus exactly the same as the continuum ones, 
but restricted to the four null directions that generate 
the reciprocal lattice.  

\subsubsection{Fermion doubling?}

Most lattice discretizations of the Dirac or Weyl equations exhibit 
{\it fermion doubling}, which means that they have zero frequency
solutions with large spatial wavevectors. As just seen,  
that does not occur for equation \eqref{onestep} if we restrict to
real frequency. However, decaying modes with zero real part of $\omega$
and small negative imaginary part would contribute to a quantity --- and thus
effectively behave as doublers --- if the decay $\exp({-|\o_I| \D t})$ is negligible over the time interval 
$\D t$ relevant to that quantity. To check for these
we need to find the complex frequency roots of the dispersion relation.

\begin{figure}
    \centering
    \vspace{-2cm}
    \begin{subfigure}[b]{0.4\textwidth}
        \includegraphics[width=\textwidth]{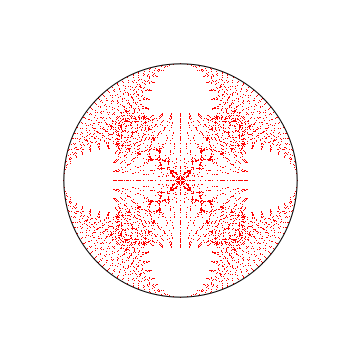}
       % \caption{All eigenvalues.}
        \label{gull}
    \end{subfigure}
    ~ %add desired spacing between images, e. g. ~, \quad, \qquad, \hfill etc. 
    \begin{subfigure}[b]{0.4\textwidth}
        \includegraphics[width=\textwidth]{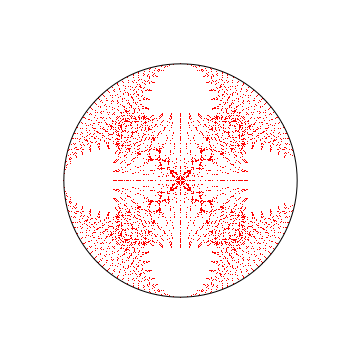}
       % \caption{}
        \label{tiger}
    \end{subfigure}
    \vspace{-1cm}
     \caption{\small Left: Eigenvalues $e^{-i\o\e}$ of $A(\theta)$, for 40 equally spaced values of each of $\theta_{1,2,3}$,
     plotted in the complex plane. 
     The black circle is the unit circle, and the real axis is horizontal. 
      Zero frequency corresponds to $1$ on the real axis, and real frequencies lie on the unit circle.
     Right: Excluding $7$ values of $\theta_1$ centered on $\theta_1=0$.}
     \label{portrait}
    \end{figure}

It seems complicated to 
develop much analytic understanding of the spectrum,
but a numerical computation easily yields the portrait in Fig.~\ref{portrait},
which depicts the eigenvalues as points on
and inside the unit circle in the complex plane.
The spectrum has fourfold rotation symmetry
since the replacement $\theta_j\rightarrow \theta_j + \pi/2$, which preserves
the condition $\sum_j\theta_j=0$ modulo $2\pi$, just multiplies the eigenvalue by 
$e^{i\pi/2}$. 
The roots $e^{-i\o\e}$ with pure negative imaginary frequency 
$\o$ lie on the positive real axis in the figure.
We learn from the figure that there is a gap in the spectrum
of these roots. Numerical experiment reveals that the maximum value on the real axis is
$1/\sqrt{3}$, which corresponds to the case $\theta_1=\theta_2=0$, $\theta_3=\theta_4=\pi$.
A solution with this eigenvalue would be damped by a factor $1/\sqrt{3}$ for each evolution 
time step. Moreoever, the figure on the right shows that if we remove a neighborhood around 
the zero wavevector, then we remove a neighborhood around zero frequency, so none of the 
solutions with nearly zero frequency correspond to large wavevectors.
We therefore expect that the lattice equation \eqref{onestep} would not exhibit 
fermion doubling phenomena in any low energy observables. This expectation is 
further corroborated in Section \ref{evaluation}, where we demonstrate that the continuum 
limit of the lattice propagator is precisely the continuum propagator.

This raises the question of how the Nielsen-Ninomiya theorem~\cite{x}
is evaded. The theorem shows that, under certain general conditions,
fermion doubling is unavoidable. One of the assumptions is that the
equation of motion is derived from a local lattice action, but on account of 
the time-asymmetric finite difference,
we did not succeed in deriving Eq.~\eqref{onestep} from any local
action without introducing further fields. 
At present, the lack of a local action is our proposed explanation
for how the fermion doubling is evaded. 

%Although there are no doubles
%in the free theory, it is possible they would arise in 
%a non-zero background gauge field, or in an interacting
%gauge theory. This possibility does not concern us here, since
%our purpose is just to construct 
%an interesting representation of the free propagator 
%as the continuum limit of
%a sum over paths on the lattice.

\section{Path integral}
Iterating Eq.~(\ref{onestep}) backwards in time, we see that
the retarded propagator between two points can be
written as a sum over paths involving
steps in the
directions ${\n{i_{1}}\dots\n{i_{N}}}$ at speed $3c$,
with the
amplitude for such a path given by the product of 
projection operators,
\be
\label{OPER}
{2^{-N}}P_{i_N}\dots P_{i_1}.
\ee
The propagator is then the sum of these operators
over all paths that connect the points.
Note that for fixed initial and final points
the number of each of the four step types is fixed, so
one sums only over the order in which the steps are taken.

In terms of the unit eigenspinor $|i\ra$ of 
$P_i=|i\ra\la i|$ the amplitude (\ref{OPER})
becomes
\be
\label{spinamp}
{2^{-N}}|i_{N}\ra \la i_N|i_{N-1}\ra\la
i_{N-1}|\cdots|i_2\ra\la i_2|i_1\ra\la i_1|.
\ee
To obtain a scalar amplitude
we may consider the matrix element of the propagator
between fixed ingoing and outgoing spinors $|i\ra$ and
$|o\ra$ selected from the set (\ref{tspinors}).
These specify directions of arrival
at the initial point and departure from the final point. 
Then, apart from the factor $2^{-N}$, the amplitude 
becomes nothing but a product of spin transition amplitudes,
\be
\label{trans}
2^{-N}\la o|i_{N}\ra \la i_N|i_{N-1}\ra\la
i_{N-1}|\cdots|i_2\ra\la i_2|i_1\ra\la i_1|i\ra.
\ee
This is a lattice version of the continuum 
spinor chain path integral of \cite{Jacobson:xt},
which involved integration over the sphere of spin states at each step, 
rather than a sum over the four states. (A cubic spatial lattice 
version was developed in \cite{TJdiss}.)

The amplitude matrix (\ref{spinamp}) is independent of the choice of phases
for the spinors, so we are free to adjust those phases to produce a
nice result if possible. Let us see what can be done.
The unit spinor corresponding to a unit vector with spherical angles
$(\theta,\phi)$ is $\bigl(\cos(\theta/2),\sin(\theta/2)\exp(i\phi)\bigr)$,
times an arbitrary overall phase. If we align the $z$ axis with one
of the tetrahedral directions, and the $x$ axis with another, then  
the spinors $|i\ra$
take the form
\bea
|4\ra&=&e^{i\delta_{4}}\left(\ba{c}1\\0\ea\right)\hfill\hspace{2cm}
|1\ra=\frac{e^{i\delta_{1}}}{\sqrt{3}}\left(\ba{c}1\\
\sqrt{2}\ea\right)\bigskip\nonumber\\
|2\ra&=&\frac{e^{i\delta_{2}}}{\sqrt{3}}\left(\ba{c}1\\
\sqrt{2}\,e^{i\frac{2\pi}{3}}\ea\right)\hfill\hspace{0.95cm}
|3\ra=\frac{e^{i\delta_{3}}}{\sqrt{3}}\left(\ba{c}1\\
\sqrt{2}\,e^{i\frac{4\pi}{3}}\ea\right),
\label{tspinors}
\eea
corresponding to the four unit vectors \eqref{COORD}.
We have identified a rather nice choice for the phases, although it requires that 
the tetrahedral symmetry be broken by singling out one preferred axis. This axis may 
be chosen to be the one determined by the ingoing spinor,  $\vec n_i=\la i|\vec\sigma|i\ra$,
and identified with the $z$ axis in the above parameterization, so that the preferred
spinor is labeled $|4\ra$. 

If we choose $\delta_4=\pi/2$ and $\delta_1=\delta_2=\delta_3=0$,
then the inner products between
the four spinors become identical up to a sign that
depends only on the order:
\be\label{PHASE}
\la1,2,3|4\ra=\la3|1\ra=\la2|3\ra=\la1|2\ra=\frac{i}{\sqrt{3}},
\ee
and the complex conjugates correspond to the opposite order. 
The $+i$ goes with bends away from the $z$ axis, or with
bends between 1, 2, and 3 that are right-handed with respect to 
the $-z$ axis.
Alternatively, we may set also $\delta_4$ to zero, so that
bends away from the $z$ axis have unit phase. For this second 
rule, the amplitude for an $N$ step path with $B$ bends and a 
net number $T$ of right-handed bends minus left-handed bends is 
\be
i^T\, {3}^{-B/2}\,2^{-N}.
\label{amplitude}
\ee
We wonder whether Feynman would have considered this beautiful and simple.

To treat left-handed Weyl spinors we need only replace the spin projections $P_i$ by the orthogonal 
projections $\bar P_i = \half(1 -\hat n_i\cdot\sigma)$, which differ just by the minus sign in place of the plus sign. 
Since the eigenspinors of $\bar P_i$ are the charge conjugates $i\sigma_y |i\ra^*$
of the four spinors (\ref{tspinors}), we conclude 
that, for left-handed spinors, $T$ in the amplitude (\ref{amplitude})
is replaced by $-T$. Note that we cannot interchange the 
right- and left-handed rules by a change of phase choice.
This would require choosing $e^{i(\d_1-\d_2)}=e^{i(\d_2-\d_3)}=e^{i(\d_3-\d_1)}=-1$.
Multiplication of these three phase factors together reveals that this is impossible.

\subsection{2+1 dimensions}

The entire discussion so far can easily be adjusted to apply in the case of 2+1 spacetime dimensions.
The Weyl equation can be adapted to 2+1 dimensions simply by omitting the $z$ direction. The resulting
equation is also the full massless Dirac equation.
Instead of the four unit step vectors we now have three, 
which can be taken to lie in the $xy$ plane separated by 
$2\pi/3$ radians, the step speed becomes $2c$, and the identity \eqref{sigma} becomes
\be
\sigma^{\mu}=\frac{2}{3}\sum_{i}P_{i}\, \N{i}.
\label{sigma2}
\ee
The spinors whose polarization vector lies in the $xy$ plane have components
$(1, e^{i\phi})/\sqrt{2}$, up to an independent overall phase for each one. Hence for the spinors corresponding
to the three steps, in the directions $\phi = 0, 2\pi/3, -2\pi/3$,
we can adopt $(1,q)/\sqrt{2}$, where $q=1, e^{i2\pi/3},e^{-i2\pi/3}$ are the cube roots of unity. The inner product of two of these is $e^{\pm i\pi/3}/2$, where the sign corresponds to clockwise or counterclockwise turns. For example, 
$\la(1, 1)|(1,e^{i2\pi/3})\ra= 1 + e^{i2\pi/3} = e^{i\pi/3}$.\\

Since the 2+1 dimensional Weyl or massless Dirac equation does not break parity symmetry,  
it should also be possible to choose the phases of the three spinors so that the amplitude rule is 
the conjugate one. Indeed, this results from the spinors $(1,1)$, $(e^{-i2\pi/3},1)$, $(e^{i2\pi/3},1)$,
all divided by $\sqrt{2}$. In addition, there is one more symmetric choice, which yields inner product 
$-1/2$ for both left and right turns! This corresponds to the choice of spinors $(1,1)$, $(e^{i2\pi/3},e^{-i2\pi/3})$, $(e^{-i2\pi/3},e^{i2\pi/3})$. All three of these rules have the same amplitude, 
proportional to $(-1)^n$, for a path that winds completely around in direction $n$ times. This last rule assigns to any path of $N$ steps the simple amplitude 
\beq
(-2)^{-B}\,2^{-N},
\eeq
where $B$ is the total number of bends.

\section{Evaluation of the path integral}
\label{evaluation}
We now evaluate the sum over paths
for the lattice propagator
${K}_{\epsilon}(\D x)$
for a spacetime displacement
$\D x$
and demonstrate
that it reproduces the continuum retarded propagator in the
limit $\epsilon\rightarrow0$.

The lattice displacement $\D x$ can be expanded
in terms of the four basis vectors:
\beq
\D x^\mu = \sum_j\D x^j \N{j},
\label{Dxi}
\eeq
hence the displacement is determined by a
unique set of four integers $N^j=\D x^j/\epsilon$.
The constraint that a path
makes a given displacement can be incorporated
as four Kronecker deltas, which we express in a Fourier
representation:
\be\prod_j\delta(N^j, {\D x^j}/{\epsilon})
=\int_{-\pi}^{\pi}
\frac{d^4\theta_j}{(2\pi)^4}
e^{i \theta_j (N^j-{\D x^j}/{\epsilon})}.
\ee
Here, and in the following, summation over a repeated $j$ index is implicit.
The lattice propagator is given by
\be
{K}_{\epsilon}(\D x) =
\sum_{N=0}^{\infty}\int_{-\pi}^{\pi}
\frac{d^4\theta_j}{(2\pi)^4}e^{-i \theta_j\D
x^j/{\epsilon}}[\amp]^{N},
\label{PROP}
\ee
where
\beq
\amp=\frac{1}{2}\sum_jP_j\, e^{i\theta_j}.
\label{A}
\eeq
The sum
over $N$ of $[\amp]^N$ produces every possible sequence of
projection operators, each with the appropriate exponential
factor encoding the number of steps in each direction.
When the integrals over $\theta_j$ are carried out,
only those step sequences that produce the displacement
$\D x^\mu$ will survive. In particular, only the
value of $N$ equal to the total number of steps
contributes.

As the step size $\epsilon$ goes to zero, the number of steps
$N$ for a fixed time interval
goes to infinity as $\Delta t/\epsilon$. Convergence
thus requires that the norm $\|\amp\|$
(i.e. the maximum norm of $\amp$ acting on a unit spinor)
be less than or equal to unity. 
It is shown in the appendix
that $\|\amp\|<1$ except when 
at least three of the $\theta_j$ coincide.
As $N$ becomes larger, $[\amp]^N$ 
therefore converges to zero pointwise except at 
this set of degenerate $\theta_j$ values, which 
has measure zero. It follows that only an $O(1/N)$
neighborhood of $\theta_j=0$ contributes 
in the $\e\rightarrow 0$ limit.

In terms of the new variables
$k_j:=\theta_j/\epsilon$, (\ref{PROP}) takes the form
\be
{K}_{\epsilon}(\D x) =
\epsilon^4\sum_{N=0}^{\infty}\int_{-\pi/\epsilon}^{\pi/\epsilon}
\frac{d^4k_j}{2\pi}e^{-i k_j\D
x^j}[\ampk]^{N}.
\label{PROPk}
\ee
In the limit $\epsilon\rightarrow0$, only an $O(1/\D t)$ neighborhood of $k_j=0$ 
contributes, so we may use the exponential approximation,
\beq
[\ampk]^N\approx e^{iN\epsilon k_j P_j/2},\label{Nepsquared}
\eeq
dropping the $O(N\epsilon^2k^2)$ correction. 
Moreover the limits of integration in (\ref{PROPk}) approach $\pm\infty$,
hence (\ref{PROPk}) yields
\be
{K}_{\epsilon\rightarrow0}(\D x) =
\epsilon^4\sum_{N=0}^{\infty}\int_{-\infty}^{\infty}
\frac{d^4k_j}{2\pi}e^{-i k_j\D x^j}
e^{iN\epsilon  k_jP_j/2}
\label{PROPk0}
\ee
To render this in more familiar terms, we next change the
variable of integration from the $k_j$ to the generic wave co-vector
components $k_\mu$. These are related via 
\beq
k_j=k_\mu n_j^\mu,
\label{ki}
\eeq
so that $k_j\D x^j = k_\mu \D x^\mu$.
Substituting (\ref{ki}) for $k_j$, and using (\ref{sigma}),
the sum in the last exponent of (\ref{PROPk0}) becomes
$k_\mu\sigma^\mu$.
The Jacobian
$|{\partial k_j}/{\partial k_\mu}|=|n_j^\mu|$ can be
computed using an explicit form of the tetrad
of unit 3-vectors, yielding
 \be
d^4k_j = 48\sqrt{3}\,d^4 k^\mu.
\ee

The final step in taking the limit is to replace the discrete variable
$N$ by a continuous one $s=N\epsilon$, in terms of which
the sum $\sum_N$ becomes $\int ds/\epsilon$. With this replacement,
and the change of variables from $k_j$ to $k_\mu$,
(\ref{PROPk0}) becomes
\be
{K}_{\epsilon\rightarrow0}(\D x) =
48\sqrt{3}\epsilon^3\int_0^\infty ds\int_{-\infty}^{\infty}
\frac{d^4k}{(2\pi)^4}e^{-ik_{\mu}\D x^{\mu}}
e^{is k_\mu\sigma^\mu}
\label{PROPs}
\ee
Except for the peculiar factor in front, this is just the
continuum retarded propagator. We can either first do the $s$ integral, 
obtaining the spacetime Fourier transform of $1/\s^\mu k_\mu$, or
first do the integral over $k_0$, which produces a Dirac
delta function $\delta(s-\D x^0)$. Then the $s$-integral
sets $s$ equal to $\D x^0$ (assuming $\D x^0>0$),
yielding the retarded propagator as
a spatial Fourier transform.
%over $s$ produces (assuming $\D x^0>0$)
%
%\be
%{K}_{\epsilon\rightarrow0}(\D x) =
%48\sqrt{3}\epsilon^3\int_{-\infty}^{\infty}
%{d^3k\over{(2\pi)^3}}e^{-ik_{a}\D x^{a}}
%e^{i\D x^0 k_a\sigma^a}.
%\label{PROP3}
%\ee
%
%This is proportional to the retarded propagator in the
%usual form of a three dimensional Fourier transform.
%Both (\ref{PROPs}) and (\ref{PROP3}) are easily
%seen to satisfy the homgeneous Weyl equation, except at
%$\D x^0=0$ where there is delta function source term.
Had we kept the subleading terms of order
$N\epsilon^2$ ($=s\epsilon$) in (\ref{Nepsquared}) the
convergence factor for the integration
limit  $s\rightarrow\infty$ would presumably have been supplied
much as in~\cite{Jacobson:xt} or \cite{Jacobson:1984bk}.

%To derive the latter directly, write the Weyl equation
%(\ref{Weyl}) in
%the form $i\\partial_t\Psi=H\Psi$, with
%the Hamiltonian $H=\vec\sigma\cdot\vec{p}$.
%The retarded propagator is just the matrix elements
%of the evolution operator $\exp(-iHt)$,
%
%\bea
%\label{PSI}\nonumber
%\Psi(\mathbf{x},t)&=&\int\d^3
%x'\langle\mathbf{x}|e^{i\mathbf{p}\cdot\vec\sigma(t-t')}|\mathbf{
%x'}\rangle\la \mathbf{x'} \Psi(t')\\
%&=&\int \d^3 x'\int_{-\infty}^\infty{\d^3
%k\over(2\pi)^3}\,e^{i\mathbf{k}\cdot\vec\sigma(t-t')}
%e^{-i\mathbf{k}\cdot(\mathbf{x}-\mathbf{x'})}\Psi(\mathbf{x'},t'),
%\eea
%
%where the second line can be obtained by inserting a complete
%set of momentum eigenstates...

It remains to account for the pre-factor $48\sqrt{3}\, \epsilon^3$.
We computed the propagator
to go between two points on the lattice. In the continuum,
the amplitude to arrive at one point starting from another
point is zero, since only by integrating
over a finite region should a nonzero amplitude arise.
The prefactor is none other
than the volume per point 
in the lattice \eqref{Vp},
with the step length 
$a$ 
equal to $3\epsilon$. 
Hence what we have actually
obtained is the continuum propagator integrated over the volume
associated with one lattice point.

\section{Mass}

So far we have discussed the theory of chiral fermions, i.e.\ fields satisfying the 
Weyl equation. 
The effect of a Dirac mass $m$ can be included
by allowing for chirality flips between right- and left-handed s
pinor propagation at each time step, 
with an associated amplitude factor 
$i\epsilon m$~\cite{Jacobson:xt}, 
as on Feynman's checkerboard. This seems to introduce new, mixed chirality
spinor transition overlaps, however those can be avoided by enlarging the lattice.
We discuss first a Dirac mass, and then a Majorana mass, in the following two subsections.

\subsection{Dirac mass}
A spin-1/2 particle with a Dirac mass is composed of two 2-component 
spinor fields, $L$ and $R$, 
which for a given spin state propagate in opposite spatial directions. Equivalently, 
for a given propagation direction they have opposite spins.
The mass term in the continuum Dirac equation couples these two fields,
\beq\label{DiracWeyl}
\s^\m \partial_\m R = im L,\qquad
\bar\s^\m \partial_\m L = imR.
\eeq
Here $\bar\sigma^\mu = (1,-\vec\sigma)$ is the set of matrices that goes with the left handed spinor. 
Together with the identity $\s^{(\m}\bar\s^{\nu)}=\eta^{\m\nu}$, 
these equations imply that each component of both spinors satisfies 
the Klein-Gordon equation, $(\square + m^2)(R,L) = 0$.

Discretization of the Dirac equation \eqref{DiracWeyl} on our lattice 
yields coupled one step evolution equations,
\bea
R(x) &=& \half\sum_i P_i \,\bigl(R+i\epsilon m\, L\bigr)(x-\e n_i)\label{Diracdiscrete1}\\
L(x) &=& \half\sum_i \bar P_i \, \bigl(L +i\epsilon m\,  R\bigr)(x-\e n_i).\label{Diracdiscrete2}
\eea
If we expand these equations in powers of $\e$, the $O(\e^0)$ term holds identically, and the $O(\e)$ terms reduce to \eqref{DiracWeyl}. To $O(\e)$, it doesn't matter if we evaluate $R$ and $L$ in the mass term at $x$ or at $x-\e n_i$, or anywhere else that differs from $x$ at $O(\e)$.

Iterating these equations yields a sum over paths, 
with an amplitude $i\e m$ to swap chiralities at each step. 
Notice that the amplitude to continue in the same direction upon a chirality change vanishes since, for each $i$,  $\bar P_i P_i = 0$. 
One can sum over the paths in spacetime, and for each path sum over the binary choices of whether
the step is an $R$ step or an $L$ step. To illustrate this consider  
for example a three-step path, ending with $R$. The corresponding contribution to the 
propagator takes the form
\begin{align}
PPP\,R &+
i\e m[PPP+PP\bar P +P\bar P\bar P]\,L \nonumber\\
&-(\e m)^2 [PP\bar P+P\bar P\bar P +P\bar PP] \,R
- i(\e m)^3P\bar P P\, L.
\end{align}
Inside a string of $\bar P$'s the bend amplitudes are just the conjugates of what they 
are inside a string of $P$'s, i.e.\ the right- and left-handed turn rules are reversed. However, 
at a transition from $P$ to $\bar P$ or vice versa, another rule is needed, because one of the four
right-handed spinors meets one of the conjugate spinors. We do not consider this beautiful and simple. 

We have found a way to improve on the elegance, but at the cost of a loss of economy in the lattice
structure and the need to restrict chiral spinors to certain paths on the lattice. It works as follows. 
Adopt now a spatial, body-centered cubic (bcc) lattice, with the distance from a cube center to a cube corner 
$3\e$, and time steps $\e$. The eight steps from each lattice point to the corners of the surrounding cube 
can be grouped into two tetrahedral sets of four. Call them the $R$ steps 
and the $L$ steps. For every $R$ step from a point there is an opposite $L$ step. Now rather than 
the $L$ chirality spinor taking the same spatial steps as the $R$ but with opposite spin, we have it taking the 
opposite steps with the same spin. This way we may work with just one set of four spin projection operators. 
The amplitude now depends on a pair of ingoing 2-component spinors, and a pair of outgoing ones. 
To express the product of projection operators in terms of bend amplitudes, as in \eqref{amplitude},
we may proceed as in the massless case, choosing, say, the incoming $R$ spinor axis as the preferred one. 
For example, an $RL$ bend from $\hat{n}_1$ to $-\hat{n}_3$ has the mass amplitude $(i\e m)$ 
and the same bend factor $i/\sqrt{3}$ as would have had an $RR$ bend from $\hat{n}_1$ to $\hat{n}_3$.

\subsection{Majorana mass}
A spin-1/2 particle with a Majorana mass has no more states than a massless, chiral particle.
Where the Dirac equation couples the opposite chirality fields, the Majorana equation couples
the field to its own orthogonal spin state.
The Majorana equation in the chiral representation is obtained by replacing $mL$ 
on the right-hand side of the Dirac equation \eqref{DiracWeyl} 
by $mCR$, where $CR:=i\s_2 R^*$ is---with a particular phase choice---the left-handed 
spinor orthogonal to $R$: $(C\psi)^\dagger R =  (i\s_2R^*)^\dagger R = -iR^T\s_2R = 0$.
Changing notation
from $R$ to $\psi$, the discrete Majorana equation can thus be chosen as
\beq\label{Majorana}
\psi(x) = \half\sum_i P_i \bigl(1+i\e m C\bigr) \psi(x-\epsilon n_i).
\eeq
In words, there is an amplitude $-i\e m$ to conjugate $\psi$ before propagating.
To first order in $\e$, we can equivalently adopt the equation
\beq\label{Majorana}
\psi(x) = \half\sum_i \bigl(1+i\e m C\bigr)P_i  \psi(x-\epsilon n_i),
\eeq
which propagates and then conjugates.

When this is iterated, we obtain a path integral in which at each step one has the amplitude $i\e m$ to 
insert the conjugation operator. For example, if there are two conjugations, the resulting path amplitude takes the form
\beq
(i\e m)^2 P\cdots PCP\cdots PCP\cdots P.
\eeq
Using the relations 
\beq\label{relns}
C^2=-1, \qquad  CP=\bar P C,\qquad PC = C\bar P.
\eeq
we can move the $C$'s to the left. When $C$ passes $P$ it turns it into $\bar P$, and when it meets another $C$ they annihilate to $-1$. Thus, for example, the above amplitude is equivalent to 
\beq
(\e m)^2 P\cdots P\bar P\cdots \bar PP\cdots P,
\eeq
where the $(-1)$ combined with the $i^2$ to give $1$. For paths with an even number of $C$'s, this
pattern will persist, each action of $C$ resulting in a chirality flip. For paths with an odd number of $C$'s, there is a remaining factor of 
$i\e m C$ at the left end. Similar to the Dirac mass case, one can trade the conjugation of the spinors for a toggling to the opposite directions for the allowed spatial steps on an enlarged lattice.

\section{Discussion}

Let us summarize what has been done in this paper. 
We discretized the Weyl equation for a massless, 
spin-1/2 particle on a time-diagonal, hypercubic spacetime lattice
with null faces. The tetrahedral light cone then just encloses the spherical, 
continuum lightcone,
ensuring fulfillment of the Courant stability condition, and the step speed is $3c$. With this 
particular step speed, the amplitude for a particle of right-handed chirality to step from a 
lattice point is $\half$ times the spin projection operator 
in the step direction, while for left-handed it is the orthogonal projector.
Iteration yields a path integral for the retarded propagator, 
with matrix path amplitude the product of 
projection operators times $2^{-N}$. 
With any particular phase choices for the 
four eigenspinors, the product of projection matrices becomes a 
product of consecutive spinor inner products,
$\la \lambda_{n+1}|\lambda_n\ra$, and the choice can be made 
so that these give rise to the 
amplitude $i^{\pm T}\, {3}^{-B/2}\,2^{-N}$ for a path 
with $N$ steps, $B$ bends, and $T$ right-handed minus left-handed bends, where the sign 
corresponds to the chirality. 
We evaluated the path integral and verified that it converges to the continuum one
in the continuum limit, and showed that fermion doubling does not occur. 
Although the underlying lattice is not Lorentz invariant, that symmetry is 
recovered by the propagator in the continuum limit.

The Dirac equation describes a pair of Weyl particles of opposite chirality, coupled by a mass term.
The mass term introduces the amplitude $i\e m$ to flip chirality in any given time step $\e$, where $m$ is the mass.
In 2D, this yields Feynman's original checkerboard path integral. 
In 4D, the chiral spinor amplitudes are punctuated by 
chirality flips. To maintain the simplest possible bend amplitudes at the chirality flips we 
enlarged the lattice to a bcc one, so that opposite chirality spinors with the same spin could step in 
opposite directions. We also showed how a Majorana mass can be similarly introduced, with an
amplitude to undergo charge conjugation. 

Our motivation here has only been to explore the relation between spin and translation, 
and to find a pleasing path integral for spinors involving just a simple rule for bend amplitudes.
Nevertheless, could it possibly be useful in a practical sense, for example in a
non-perturbative, lattice computation scheme  for chiral gauge theories?
The first step in that direction is easy. 
An external electromagnetic or non-abelian gauge field $A$ can be included simply by 
multiplying the one-step amplitude by $P\exp(i\int A)$, where the path-ordered 
integral is taken over the spacetime step. (Something like this is standard in lattice gauge theory
computations, but occurs there with the finite differences in the action.) 
But this is where the easy answers end. Lattice gauge theory adopts Euclidean signature
so that the vacuum state or its excitations can be straightforwardly selected,
and the path integral over gauge fields can be evaluated by statistical methods. 
Another essential reason is that the symmetry of the 4D lattice theory can be sufficient to 
ensure 4D rotation invariance in the continuum limit. By contrast,  Lorentzian 
lattice schemes ensuring 4D Lorentz symmetry in the continuum limit of an interacting 
theory are not known. This, together with the fact that our path integral constructs the retarded 
propagator rather than the Feynman propagator, makes any directly useful application 
in its present form seem unlikely.

As a model of relativistic quantum propagation in 
a discrete spacetime, our scheme has a serious flaw: 
the discrete propagator is not unitarity.
This is not because discreteness and unitarity are necessarily in conflict.
Indeed, as shown in \cite{Bialynicki-Birula:hi},
one can write a unitary discrete
evolution rule on a body centered cubic lattice  whose continuum limit is the Weyl equation.
(Interestingly, unitarity and locality were shown there to {\it imply} the
Weyl equation.) 
But consider an initial state that is non-vanishing only at one lattice point, with
normalized spin state $|\psi\ra$.
At the next time step, according to (\ref{onestep}),
it has support at the four corners of a tetrahedron, with the amplitudes
$\frac{1}{2}P_i|\psi\ra$. The norm of the state after one step is then
$\la\psi|\sum_i \frac{1}{4}P_i|\psi\ra=\frac{1}{2}$, i.e. it has
decreased by a factor of two, violating unitarity.
Also, the evolutions of orthogonal states
are not orthogonal. 
Two points at one time
have either one or zero common points in their one-step future.
In the former case, the one-step evolutions of two orthogonal
states supported
on the two initial points are clearly not orthogonal,
because they overlap in just one point which will
make the unique non-zero (since $P_iP_j\ne0$)
contribution to the inner product.
%This means that initially disjoint 
The evolutions
therefore
have ``more overlap than they should", which
presumably counteracts the loss of norm of each individual evolution in
such a way that the continuum limit is unitary.

To conclude, what looks like an innocent game of checkers has a deep and 
not so hidden connection with a more serious agenda. 
In print and in person, 
David Finkelstein credited conversations with Feynman for inspiring the notion, much explored in 
David's work, that spin is the ``growing tip" of spacetime paths. 
The idea is that the Dirac spin matrices $\gamma^\mu$ somehow describe quantized 
displacements. One can find the seed of this idea in Feynman's ``Operator calculus" 
paper \cite{Feynman:1951gn}. 
Feynman invents the device of attaching a continuous index to operators, which keeps track 
of their order of multiplication. In Appendix B of his paper, he uses this method 
to derive a formal expression for the 
electron self-energy, in which the derivative operator is eliminated and the role of the
position is transferred to an integral, $\int_0^w \gamma^\mu(s) ds$, over order-indexed gamma matrices. 
Feynman then remarks that ``all reference 
to space coordinates have disappeared" from the expression, and he adds, parenthetically,
%
%\begin{quote} 
``It is suggestive that perhaps coordinates and the space-time they represent may in some future theory 
be replaced completely by an analysis of ordered quantities in some hypercomplex algebra".
%\end{quote}
David sought that future theory.

\section*{Acknowledgements}
We are grateful to E.~Hawkins for
mathematical aid, and to Yigal Shamir and Paulo Bedaque for instruction about lattice fermions.
This work was supported in part by the NSF under grants
PHY-9800967, PHY-0300710, PHY-0601800, PHY-0903572, PHY-1407744	
at the University of Maryland, in part by the CNRS at the
Insitut d'Astrophysique de Paris, and by 
Perimeter Institute for Theoretical Physics.  
Research at Perimeter Institute is supported by the 
Government of Canada through Industry Canada and by the 
Province of Ontario through the Ministry of Research \& Innovation.

\appendix
\section{Norm of the amplification matrix}

In this appendix we prove that the norm of the matrix $A(\theta)$ defined in 
Eq. (\ref{A}) is less than unity unless at least three of the
$\theta_i$ coincide, in which case the norm is unity. The proof
is due to Eli Hawkins.

Let $|\nu\ra$ be any unit spinor. The squared norm of $A|\nu\ra$ is 
$\|A|\nu\ra\|^2=\la\nu|A^\dagger A|\nu\ra$, whose maximum 
is the larger eigenvalue of $A^\dag A$.  
This value defines the squared norm $\|A\|^2$.

Using the definition of the spin projection operators
(\ref{Projector}) and the inner products of the unit vectors
$\hat{n}_i$ ($1$ if $i=j$ and $-\frac13$ if $i\ne j$) we find 
\beq
{\rm tr} (A^\dagger A) = 1 + \tfrac16 \sum_{(ij)} \cos(\theta_i-\theta_j)
\eeq
where the sum is over the 6 choices of $\{i,j\}\subset \{1,2,3,4\}$. 
This trace is at most $2$, and therefore the smaller eigenvalue of 
$A^\dag A$ is less than $1$ unless $A^\dag A=1$.
Hence
\beq
\Phi := \det \left(A^\dag A -1\right)
\eeq
has the same sign as $1 - \| {A}\| $.

The matrix $A^\dag A$ is linear in terms of
$e^{i(\theta_i-\theta_j)}$, therefore $\Phi$ is quadratic. Because
$\Phi$ is invariant under all permutations of the $\theta$'s, it can
be written as a quadratic function of the cosines
$\cos(\theta_i-\theta_j)$. Because $\Phi$ vanishes when the $\theta$'s
are all equal, it is convenient to write it in terms of the cosines
minus 1. It thus takes the form,
\bea
\label{Phi1}
\Phi &=& a \sum_{(ij)} \left(1 - \cos[\theta_i-\theta_j]\right) 
+ b \sum_{(ij)} \left(1 - \cos[\theta_i-\theta_j]\right)^2 \nonumber\\ 
&&+  c\!\!\sum_{(ij)(kl)} \left(1 - \cos[\theta_i-\theta_j]\right)
\left(1 - \cos[\theta_k-\theta_l]\right) .
\eea
The last sum is over the 3 partitions of $\{1,2,3,4\}$ into pairs. 
When $\theta_2=\theta_3=\theta_4$,  $A$ takes the form 
\eqref{case}, which obviously has an eigenvalue of unit modulus. 
Therefore $\Phi=0$ in this case, which shows that $a=b=0$.
To determine the value of $c$, consider the case that 
$\theta_1=\theta_2=0$ and $\theta_3=\theta_4=\pi$. 
Then $A=\frac12(\hat{n}_1+\hat{n}_2)\cdot\vec{\sigma}$,
so $A^\dagger A = 1/3$, hence $\Phi=4/9$. 
The last sum in \eqref{Phi1} is $8$, so $c=1/18$.

The determinant $\Phi$ is thus given by 
\beq
\label{Phi2}
\Phi = \tfrac1{18}\sum_{(ij)(kl)} 
\left(1 - \cos[\theta_i-\theta_j]\right)
\left(1 - \cos[\theta_k-\theta_l]\right).
\eeq
This satisfies $\Phi\geq 0$, therefore $\|A\|\leq 1$. Each term is
non-negative, therefore $\Phi=0$ only if every term vanishes. 
This occurs only if at least three of the $\theta$'s are equal.

\end{document}